\documentclass[prd,nofootinbib,floats,aps,twocolumn,tightenlines,runinaddress]{revtex4-2}
\usepackage{natbib}
\usepackage{graphicx}
\usepackage{mathtools}
\usepackage[normalem]{ulem}
\usepackage{hyperref}
\hypersetup{
	colorlinks=true,
	linkcolor=blue,
	filecolor=magenta,      
	urlcolor=cyan,
}
\usepackage{commath}
\usepackage{bm}
\usepackage{amsfonts,amsmath,amssymb,amsthm}
\usepackage[usenames,dvipsnames]{xcolor}
\usepackage{dsfont}
\usepackage{enumitem}
\usepackage{color}
\usepackage{tikz}
\usepackage{physics}
\usepackage{verbatim}

\newcommand{\ii}{\mathrm{i}}
\newcommand{\proj}[2]{\left| {#1} \right\rangle\!\!\left\langle {#2} \right|}

\renewcommand*\d[2][]{%
	\mathrm{d}%
	\ifx\relax#1\relax\else
	\rule{-0.02em}{1.5ex}^{#1}\rule{0.08em}{0ex}\!
	\fi
	#2\,
}

\newcommand{\rhoh}{\hat{\rho}}

\newcommand\restr[2]{{
		\left.\kern-\nulldelimiterspace 
		#1 
		\vphantom{\normal|} 
		\right|_{#2} 
}}

\newcommand{\diff}{\mathrm{d}}

\newcommand{\beq}{\begin{equation}}
	\newcommand{\eeq}{\end{equation}}

\newtheorem*{claim*}{Claim}

\begin{document}

\title{Thermodynamic bound on quantum state discrimination}
	
\author{Jos\'{e} Polo-G\'{o}mez}
\email{jpologomez@uwaterloo.ca}
\affiliation{Department of Applied Mathematics, University of Waterloo, Waterloo, Ontario, N2L 3G1, Canada\\
Institute for Quantum Computing, University of Waterloo, Waterloo, Ontario, N2L 3G1, Canada\\
Perimeter Institute for Theoretical Physics, Waterloo, Ontario, N2L 2Y5, Canada}
	
\begin{abstract}
		
We show that the second law of thermodynamics poses a restriction on how well we can discriminate between quantum states. By examining an ideal gas with a quantum internal degree of freedom undergoing a cycle based on a proposal by Asher Peres, we establish a non-trivial upper bound on the attainable accuracy of quantum state discrimination. This thermodynamic bound, which relies solely on the linearity of quantum mechanics and the constraint of no work extraction, matches Holevo's bound on accessible information, but is looser than the Holevo-Helstrom bound. The result gives more evidence on the disagreement between thermodynamic entropy and von Neumann entropy, and places potential limitations on proposals beyond quantum mechanics.
		
\end{abstract}
	
\maketitle
	
\section{Introduction}\label{Section: introduction}
	
\vspace{-0.1cm}

The major role that information plays in thermodynamics was first realized by Szilard~\cite{Szilard}, and later underpinned by Landauer's erasure principle~\cite{Landauer} and Bennett's exorcization of Maxwell's demon~\cite{Bennett}. The interplay between the second law of thermodynamics and information processing has been ever since the subject of extensive research (see~\cite{Maruyama2009} for a review). In the last decades, this relationship has been particularly fruitful for the study of small far-from-equilibrium systems~\cite{Sagawa2012smallsystems}, both in the realm of stochastic thermodynamics (see, e.g.,~\cite{Piechocinska2000,Touchette2000,Cao2004,Kawai2007,Cao2009,Abreu2011,Abreu2012,Strasberg2013,Mandal2013,Barato2014,Horowitz2014}), and quantum mechanics (see~\cite{Alicki2004,Sagawa2009,Jacobs2009,Sagawa2010,Hilt2011,Esposito2011,Sagawa2012fluctuation,Reeb2014,Alhambra2016,Naghiloo2018,Ptaszynsky2019}, among many others).  
	
The potential applications of quantum thermodynamics~\cite{Scovil1959,Sekimoto2000,Sato2002,Scully2002,Quan2007,Dillenscheneider2009,Linden2010,Zhang2014,Rossnagel2014,Huang2014,Abah2014,Quan2014,Gardas2015,Rossnagel2016,Klaers2017,Niedenzu2018,PozasKerstjens2018,Micadei2019,YungerHalpern2022} have been one of the main driving forces for its development in the last decades~\cite{Vinjanampathy2016,Binder2018,Deffner2019}, and it is precisely in this context that it is important to understand the constraints that thermodynamics imposes on quantum information processing. 
One of the most elemental problems to be considered in the theory of quantum information is quantum state discrimination, i.e., the problem of determining how good we can possibly get at distinguishing a set of quantum states from one another. 
Quantum state discrimination is essential to retrieving classical information from a quantum system, making it a crucial component in most quantum information processes~\cite{Bae2015}. It has also been shown to be relevant for quantum foundations~\cite{Gisin1998,Simon2001,Gallego2010,Pusey2012} and quantum communication~\cite{Bruss1998,Keyl1999,Bruss2000,Bae2006,Chiribella2006}.
	
That the second law of thermodynamics imposes a restriction on our ability to distinguish quantum states was first realized by Asher Peres~\cite{Peres1993}. In his work, Peres considered an ideal gas of particles with spin, and imagined the existence of two ``magical membranes'', \textit{perfectly} transparent or opaque to two non-orthogonal states. It was shown that these fictitious artifacts would allow the performance of a cycle in which heat was extracted from a heat bath and completely converted into work, in contradiction with the Kelvin-Planck statement of the second law. This result was not a challenge to the second law, of course, since the membranes imagined by Peres, to the best of our knowledge, do not exist. However, it shows that the existence of these membranes is \textit{not only} forbidden by quantum mechanics, but by thermodynamics \textit{as well}, yet establishing another link between the already very intertwined areas of thermodynamics and information. Moreover, a cycle akin to that of Peres was also considered by K. Maruyama, \v{C}. Brukner and V. Vedral~\cite{Maruyama2005} to obtain a bound for accessible information from thermodynamic arguments. This bound was similar but weaker than that obtained by Holevo~\cite{Holevo1973} within quantum information theory. 
	
In the spirit of these previous works, here we show that the second law of thermodynamics imposes a bound on how well we can distinguish two pure quantum states. In Sec.~\ref{Section: Peres' demon}, we reformulate Peres' idea introducing a ``demon'' that carries out the (in principle, \textit{forbidden}) operations, since this artifact highlights the informational aspects that will turn out to be relevant afterwards. In Sec.~\ref{Section: Modified cycle and bound}, we will obtain the ``thermodynamic bound'' on quantum state discrimination using an adequate modification of the previous setup. In Sec.~\ref{Section: Discussion} we compare this bound to the other bounds from quantum information theory and analyze the consequences of the results, concluding in Sec.~\ref{Section: Conclusion}. 
	
\vspace{-0.4cm}

\section{Peres' demon}\label{Section: Peres' demon}

\vspace{-0.1cm}
	
In this section we will reformulate the cycle devised by Peres~\cite{Peres1993,Maruyama2009} in a way that will allow us to monitor both the flow of information and the role of measurements in the process. The key elements of the reformulation are what we will call \textit{Peres' demons}. We will assume that these entities have somehow the ability to perfectly distinguish two specific non-orthogonal quantum states. Notice that, in principle, these demons are not Maxwell's demons, in the sense that the \textit{forbidden} operation that we will assume they have the ability to perform does not explicitly violate a thermodynamic law, but rather a quantum mechanical one. We will see that, nevertheless, this ability leads to the possibility of violating the second law of thermodynamics. Specifically, with the sole premise of the linearity of quantum mechanics, the second law of thermodynamics enforces that non-orthogonal states cannot be perfectly distinguished. 
	
\begin{figure*}
\begin{center}
\includegraphics[scale=1.05]{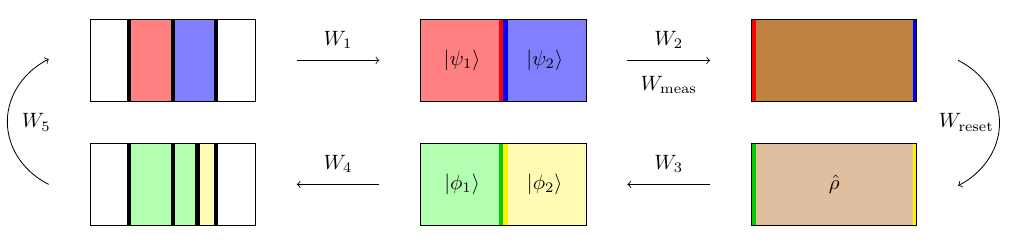}
\caption{Reformulation of the cycle proposed by Asher Peres~\cite{Peres1993}, in which assuming the ability to distinguish two different non-orthogonal quantum states allows the extraction of energy from a heat bath, and its full conversion into work, in violation of the second law of thermodynamics. States $\ket{\psi_1}$, $\ket{\psi_2}$, $\ket{\phi_1}$, and $\ket{\phi_2}$ are depicted in light red, light blue, light yellow, and light green, respectively. The Peres' demon operating the red wall lets particles in state $\ket{\psi_1}$ go through, and rejects those in state $\ket{\psi_2}$, while the one operating in the blue wall does the opposite. Similarly, the yellow membrane is transparent to $\ket{\phi_1}$ and opaque to $\ket{\phi_2}$, while the opposite is true for the green membrane. The mixed post-measurement state (with particles in states $\ket{\psi'_1}$ and $\ket{\psi'_2}$) and post-resetting state (with particles in states $\ket{\psi_1}$ and $\ket{\psi_2}$) are depicted in dark and light brown. }
\label{fig: Peres cycle}
\end{center}
\end{figure*}

We consider an ideal gas contained in a cylinder of fixed total volume $V$. The gas has an internal quantum degree of freedom that is decoupled from the rest. The cycle, depicted in Fig.~\ref{fig: Peres cycle}, starts off with a gas divided in two portions of $N/2$ particles whose internal degrees of freedom are in two \textit{non-orthogonal} states $\ket{\psi_1}$ (left, in red) and $\ket{\psi_2}$ (right, in blue). Each gas occupies a quarter of the total volume available, and is separated by a fixed opaque wall, as represented in the top left picture of Fig.~\ref{fig: Peres cycle}. 
	
In the first step, the two gases expand isothermically, allowing the extraction of work. Following the convention by which work is positive when it is given to the system, and negative when extracted from it, we have that the work involved in this first step of the process is
\begin{equation}\label{eq: W1}
W_1= - N k_{\textsc{b}} T \int_{V/4}^{V/2} \frac{\textrm{d}v}{v}=- N k_{\textsc{b}}T \ln2.
\end{equation}
	
In the second step, the opaque wall separating the two gases is exchanged by two other walls, initially at the centre of the cylinder, and one next to the other. These walls are each operated by one Peres' demon, an entity capable of perfectly distinguishing $\ket{\psi_1}$ from $\ket{\psi_2}$.  We will assume that the demons share a common memory in which they store the information about what is the internal quantum state of each particle in the gas. The job of these demons is to use their ability to distinguish $\ket{\psi_1}$ and $\ket{\psi_2}$ to decide whether a particle arriving to the wall goes through it or not, depending on what the quantum state of the particle is. The demon operating the wall to the left (red wall, in Fig.~\ref{fig: Peres cycle}) is instructed to let the particles in state $\ket{\psi_1}$ pass, rebounding off the ones in state $\ket{\psi_2}$. That way, this wall is \textit{transparent} to $\ket{\psi_1}$, while \textit{opaque} to $\ket{\psi_2}$. The wall to the right (blue wall, in Fig.~\ref{fig: Peres cycle}) operates in reverse: it is transparent to $\ket{\psi_2}$, and opaque to $\ket{\psi_1}$. Thus, the gas in the left half of the cylinder (in state $\ket{\psi_1}$) exerts pressure on the blue wall, but not the red one. Conversely, the gas in the right half (in state $\ket{\psi_2}$) exerts pressure on the red wall, but not the blue one. Thus, once the walls are allowed to move, the red and the blue walls move to their left and right, respectively, and they do so without opposition until they reach the ends of the cylinder. Each gas is therefore freely expanding to the whole volume, and the work involved in this process is
\begin{equation}\label{eq: W2}
W_2=-N k_\textsc{b} T \int_{V/2}^{V} \frac{\textrm{d}v}{v}=- N k_{\textsc{b}}T \ln2.
\end{equation}
At the end of this step, the demons have measured the state of all the particles, and this information is stored in their shared memory. The quantum state of the particles might have been modified by the measurement process as well, and the measurement itself had an associated work cost that we denote as $W_{\text{meas}}$. Now, notice that the information stored in the memory was not depleted in the performance of the second step. Thus, the demons can be instructed to perform a joint \textit{resetting} operation on the memory and the gas to reverse the measurement process, taking both systems back to their initial states, with an associated work cost $W_{\text{reset}}$. After this step, the information in the memory is erased, and the quantum state of a random particle in the gas (i.e., for the ensemble) can be described by the mixed state
\begin{equation}\label{eq: rho}
\hat\rho=\frac{1}{2}\ket{\psi_1}\!\!\bra{\psi_1} + \frac{1}{2} \ket{\psi_2}\!\!\bra{\psi_2}.
\end{equation}
Here, we will work in the limit in which the measurement-resetting process is reversible, i.e., when 
\begin{equation}
W_{\text{meas}}+W_{\text{reset}}=0.
\end{equation}
The assumption that the measurement and resetting processes can be performed reversibly might seem counterintuitive since 1) especially in quantum mechanics, measurements are regarded as highly irreversible~\cite{vonNeumann,Peres1980}, and 2) in this protocol the demons use the outcomes of the measurements to apply a feedback to the gas (see, e.g.,~\cite{Jacobs2009,Sagawa2010,Funo2013}). However, in this setup the feedback is applied to the kinematic degrees of freedom, which are dynamically decoupled from the quantum degree of freedom that was measured. This is the reason why the information in the memory is not exhausted by the feedback in the first place. As a consequence, we can think of the measurement and resetting processes as being reversible overall~\cite{Peres1974}, as long as they are only concerned with the memory and the internal quantum state. In this sense, notice that the quantum state of the gas after the resetting is the mixture given in Eq.~\eqref{eq: rho} that also corresponds to the state of the gas before step 2 \textit{if we trace out} the kinematic degrees of freedom, i.e., if we do not take into account whether the particles of the gas are to the left or to the right of the central wall (cf. Fig.~\ref{fig: Peres cycle}). It is not the state of the gas as a whole that is reversed during the resetting operation, but rather just the state of its quantum degree of freedom (jointly with the memory). The point can still be made that, e.g., projective measurements in quantum mechanics are irreversible. However, by virtue of Naimark's dilation theorem~\cite{Watrous2018}, this irreversibility can be understood as emerging from tracing out external degrees of freedom, specifically those of the measurement apparatus~\cite{Peres1974}. Even without this consideration at hand, it has been shown that sharp measurements can be approximated arbitrarily well with logically reversible measurements~\cite{Ueda1996}. It is worth remarking that these considerations were not made in Peres' work~\cite{Peres1993}, since in his version of the cycle the role of demons is played by membranes (see Sec.~\ref{Section: introduction}), and these are assumed to perfectly discriminate between $\ket{\psi_1}$ and $\ket{\psi_2}$ without altering the state of the particles, so that it is not necessary to consider an explicit measurement process---nor a memory to store its outcomes.


Now, to proceed with the rest of the cycle, notice that, in the quantum state given in Eq.~\eqref{eq: rho}, $\ket{\psi_1}$ and $\ket{\psi_2}$ are non-orthogonal, and therefore $\hat\rho$ can be diagonalized to
\begin{equation}\label{eq: rho diagonalization}
\hat\rho=c \ket{\phi_1}\!\!\bra{\phi_1} + (1-c) \ket{\phi_2}\!\!\bra{\phi_2},
\end{equation}
for some orthogonal states $\ket{\phi_1}$ and $\ket{\phi_2}$, and some real constant \mbox{$c \in [0,1]$}. We can now introduce two membranes such that one is transparent to $\ket{\phi_1}$ and opaque to $\ket{\phi_2}$ (the green one, in Fig.~\ref{fig: Peres cycle}), and the other one is opaque to $\ket{\phi_1}$ and transparent to $\ket{\phi_2}$ (the yellow one, in Fig.~\ref{fig: Peres cycle}). Since $\ket{\phi_1}$ and $\ket{\phi_2}$ are orthogonal, these membranes are in principle physically realizable (even though they are highly ideal~\cite{Peres1993}).
	
In the third step, each membrane is introduced at one different end of the cylinder to compress isothermically the portion of the gas they are opaque to, until they meet at the centre of the cylinder. The work involved in this process is
\begin{equation}\label{eq: W3}
W_3=-[cN +(1-c)N]k_{\textsc{b}} T \int_{V}^{V/2} \frac{\diff v}{v}=N k_{\textsc{b}} T \ln 2, 
\end{equation}    
and as a result the gas is separated in two portions, one in state $\ket{\phi_1}$ (light green in Fig.~\ref{fig: Peres cycle}), and the other one in state $\ket{\phi_2}$ (light yellow in Fig.~\ref{fig: Peres cycle}). 

In the fourth step of the process, each portion of the gas is isothermically compressed to the initial pressure, namely,
\begin{equation}
P_0= \frac{\frac{N}{2} k_{\textsc{b}}T}{\frac{V}{4}}=\frac{2Nk_{\textsc{b}}T}{V}.
\end{equation}
This compression is carried out using regular opaque walls, depicted in black in Fig.~\ref{fig: Peres cycle}. Since the portion of gas in state $\ket{\phi_1}$ contains $cN$ particles, it has to be compressed to a volume
\begin{equation}
V_1=\frac{cN k_{\textsc{b}}T}{P_0}=\frac{cV}{2}.
\end{equation}
Analogously, the portion of gas in state $\ket{\phi_2}$ has to be compressed to a volume
\begin{equation}
V_2=\frac{(1-c)V}{2}.
\end{equation}
The work involved in this process is
\begin{align}\label{eq: W4}
W_4&=-cN k_{\textsc{b}} T \int_{V/2}^{cV/2} \frac{\diff v}{v} - (1-c)N k_{\textsc{b}} T \int_{V/2}^{(1-c)V/2} \frac{\diff v}{v} \nonumber \\
&= N k_{\textsc{b}} T [-c \ln c -(1-c)\ln (1-c)] \nonumber \\
&=N k_{\textsc{b}} T S(\hat\rho) \ln 2,
\end{align}
where we have recognized that
\begin{equation}
-c \log_2 c -(1-c)\log_2 (1-c)=H(c)=S(\hat\rho)
\end{equation}
is the Shannon entropy of the binary distribution with probability $c$, $H(c)$, and that this is precisely the von Neumann entropy of the state $\hat\rho$, since $c$ and $1-c$ are its eigenvalues. Note that, as long as $\ket{\psi_1}$ and $\ket{\psi_2}$ are non-orthogonal, we have that $c \neq 1/2$, and therefore \mbox{$H(c) < H(1/2)=1$} (see Appendix~\ref{Appendix: Even mixture of non-orthogonal states}).
	
Finally, in the fifth and last step of the cycle, an additional wall is introduced in the cylinder. This leaves the gas divided in three portions. To explain this last process we will refer to the bottom left picture of Fig.~\ref{fig: Peres cycle}, where we have assumed, without loss of generality, that $c>1/2$, and hence there are more particles of the gas in state $\ket{\phi_1}$. In that case, two portions of gas, occupying volumes $V/4$ and $(2c-1)V/4$, are in state $\ket{\phi_1}$, while the remaining portion, occupying a volume $(1-c)V/2$, is in state $\ket{\phi_2}$. We can then perform separate unitary transformations in each portion. On the leftmost one (occupying a volume $V/4$), we perform a unitary taking $\ket{\phi_1}$ to $\ket{\psi_1}$. Meanwhile, on the central and rightmost portions (whose volumes add up to $V/4$), we perform two different unitaries that transform $\ket{\phi_1}$ and $\ket{\phi_2}$ into $\ket{\psi_2}$, respectively. These unitaries---which, for instance, in the case in which the internal quantum degree of freedom is spin, could correspond to the application of magnetic fields with a certain direction, intensity, and duration---can be performed with an arbitrarily small work cost, which becomes zero in the quasistatic limit, when we allow the process to take place during an infinitely long period of time. Thus, we can assume
\begin{equation}\label{eq: W5}
W_5=0.
\end{equation} 

Once the cycle is completed, we can compute the total work involved:
\begin{align}
W_t&=\sum_{j=1}^{5} W_j + W_{\text{meas}} + W_{\textrm{reset}} \\
&=-N k_{\textsc{b}} T [1-S(\hat\rho)] \ln 2. \nonumber
\end{align}
Since, as remarked before, $S(\hat\rho)<1$ as long as $\ket{\psi_1}$ and $\ket{\psi_2}$ are not orthogonal, we conclude that
\begin{equation}
W_t < 0,
\end{equation}
which means that the cycle's net effect is the extraction of work, in violation of the second law. This, as anticipated, is a consequence of having assumed the possibility of distinguishing two non-orthogonal states (while still keeping the well-known ability to discriminate orthogonal states~\cite{vonNeumann}), which turns out to be in conflict with the second law.

\section{Thermodynamic bound}\label{Section: Modified cycle and bound} 
	
In this section we modify the setup presented in Sec.~\ref{Section: Peres' demon} to obtain a bound for how well we can discriminate two quantum states solely based on thermodynamic constraints. To do so, instead of Peres' demons, we consider demons that can only distinguish states $\ket{\psi_1}$ and $\ket{\psi_2}$ with a certain probability of success. We then calculate the total work involved in the cycle with these conditions. Imposing that the second law is satisfied we obtain an upper bound on the accuracy with which a pair of second-law-abiding demons should be able to discriminate $\ket{\psi_1}$ from $\ket{\psi_2}$. Notice that, unlike before, these entities are given the ability to distinguish quantum states \textit{only within the limits of thermodynamics}. In Sec.~\ref{Section: Peres' demon}, Peres' demons were considered to have the ability to perfectly distinguish a pair of non-orthogonal quantum states, which is in principle an operation forbidden by quantum mechanics, but turned out to be forbidden by thermodynamics as well. Here, the second-law-abiding demons have the ability to perform the discrimination of non-orthogonal quantum states only with a certain efficiency limited by the fulfillment of the second law of thermodynamics. We should nevertheless still call them demons since, as long as the thermodynamic bound is looser than the quantum information one---as we will see is indeed the case---, these entities can perform operations that are forbidden by quantum mechanics (but might not be forbidden by other proposals beyond standard quantum theory). 
	
\begin{figure*}
\begin{center}
\includegraphics[scale=1.05]{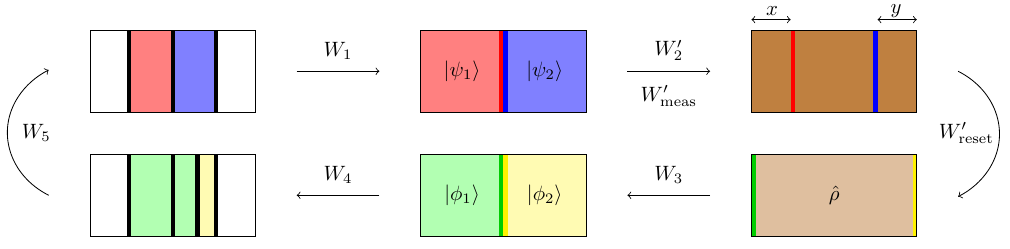}
\caption{Modified cycle resulting from having demons that cannot \textit{perfectly} distinguish the two non-orthogonal states $\ket{\psi_1}$ and $\ket{\psi_2}$. States $\ket{\psi_1}$, $\ket{\psi_2}$, $\ket{\phi_1}$, and $\ket{\phi_2}$ are depicted in light red, light blue, light yellow, and light green, respectively. The demon operating the red wall lets particles \textit{that it measures to be} in state $\ket{\psi_1}$ go through, and rejects those measured to be in state $\ket{\psi_2}$, while the one operating in the blue wall does the opposite. On the other hand, the physically realizable yellow membrane is transparent to $\ket{\phi_1}$ and opaque to $\ket{\phi_2}$, while the opposite is true for the green membrane.}
\label{fig: modified cycle}
\end{center}
\end{figure*}

The modified cycle is represented in Fig.~\ref{fig: modified cycle}. There, it becomes apparent that steps 1, 3, 4, and 5 of the cycle are the same as in Sec.~\ref{Section: Peres' demon}, and so is the work involved in them. The only differences we ought to analyze are in step 2 and the resetting of the demons' memory.
	
In this new setup, the demons are only capable of distinguishing $\ket{\psi_1}$ and $\ket{\psi_2}$ with some limited efficiency: namely, we assume that for some $\delta \in [0,1]$,
\begin{equation}\label{eq: prob. success}
p_s=\operatorname{Prob}(\textrm{``guessing correctly''})=\frac{1+\delta}{2}.
\end{equation}
Note that $\delta=0$ corresponds to the case in which the demons are not capable of discriminating one state from the other at all, while $\delta=1$ corresponds to the scenario studied in Sec.~\ref{Section: Peres' demon}. 
	
In the second step of the cycle, the demons operate the red and blue walls imperfectly, and unlike in the previous setup, the process will in general stop before the walls reach the ends of the cylinder. Specifically, a fraction \mbox{$(1-\delta)/2$} of the particles that were in state $\ket{\psi_1}$ are mistakenly identified as being in state $\ket{\psi_2}$. These particles now exert pressure on the red wall, and are constrained to move between it and the leftmost end of the cylinder. Similarly, a fraction $(1-\delta)/2$ of the particles that were in state $\ket{\psi_2}$ are mistakenly identified as being in state $\ket{\psi_1}$, and they exert pressure on the blue wall, constrained to move between it and the rightmost end of the cylinder (see top right picture in Fig.~\ref{fig: modified cycle}, for clarity). Thus, the red wall feels pressure 
\begin{itemize}
\item[-] on its left, from those particles initially in state $\ket{\psi_1}$ that were measured to be in state $\ket{\psi_2}$, and
\item[-] on its right, from those particles that were correctly measured to be in state $\ket{\psi_2}$, which move freely between the red wall and the rightmost end of the cylinder (the blue wall is transparent to them, since the demons share the memory, i.e., their measurements are consistent). 
\end{itemize}
Since the red wall is transparent to the rest of particles, it stops moving whenever these two pressures equalize. Let $x$ be the fraction of the total volume to the left of the red wall once it reaches equilibrium (cf. Fig.~\ref{fig: modified cycle}), then, by Eq.~\eqref{eq: prob. success}, 
\begin{equation}
\frac{1-\delta}{2} \,\frac{NT}{2xV}=\frac{1+\delta}{2} \, \frac{NT}{2(1-x)V} \; \Rightarrow \; x=\frac{1-\delta}{2}.
\end{equation}
An analogous treatment for the blue wall yields
\begin{equation}
y=\frac{1-\delta}{2},
\end{equation}
where now $y$ is the fraction of the total volume to the right of the blue wall once it reaches equilibrium (cf. Fig.~\ref{fig: modified cycle}). The work involved in these ``incomplete expansions'' is then
\begin{align}\label{Eq: W'2}
&W'_2=-\frac{1-\delta}{4} N k_{\textsc{b}} T \int_{V/2}^{xV} \frac{\diff v}{v} - \frac{1+\delta}{4} N k_{\textsc{b}} T \int_{V/2}^{(1-x)V} \frac{\diff v}{v} \nonumber \\
&\phantom{==\;\;\;} - \frac{1-\delta}{4} N k_{\textsc{b}} T \int_{V/2}^{yV} \frac{\diff v}{v} - \frac{1+\delta}{4} N k_{\textsc{b}} T \int_{V/2}^{(1-y)V} \frac{\diff v}{v} \nonumber \\
&=\!-Nk_{\textsc{b}} T \ln 2 \, \bigg( 1 + \frac{1-\delta}{2}\log_2 \frac{1-\delta}{2} + \frac{1+\delta}{2} \log_2\frac{1+\delta}{2} \bigg) \nonumber \\
&=\!-N k_{\textsc{b}} T \ln 2 \, \bigg[ 1-H \bigg( \frac{1+\delta}{2}\bigg) \bigg],
\end{align}
where $H(p)$ is again the Shannon entropy of the binary distribution with probability $p$. Notice in particular that when $\delta=1$, $H(1)=0$ and we recover the result in Eq.~\eqref{eq: W2}. 
	
It is worth remarking that, as before, the measurements, with an associated work cost $W'_{\text{meas}}$, might have affected the state of the particles. One might wonder whether the post-measurement states of the particles \textit{measured to be} in state $\ket{\psi_k}$ ($k=1,2$) are all the same or, on the contrary, they are different depending on their initial state. We argue here that, for the process to be optimal (and so we need it to be if we want to extract an upper bound for $\delta$), these states have to be the same. If they were not, then they would be physically distinguishable to some extent, and additional measurements could be performed by the demons to refine their initial guessing (i.e., there would be still some information left in the post-measurement states that could be extracted by the demons). Thus, we can assume here that all particles measured to be in state $\ket{\psi_k}$ (whether correctly or incorrectly) are left in the same post-measurement state $\ket{\psi'_k}$. Another consequence of this is that, once a particle is measured, it is unequivocally identified with the state of its measurement outcome by both demons (since they share their memory), and it will behave accordingly with respect to both walls. This condition is tantamount to requiring that the outcomes of sequential measurements performed by the demons have to be consistent. 
	
Moreover, with the previous assumption, the mixture of gases at the end of step 2 turns out to be homogeneous throughout all the cylinder. Take, for instance, the leftmost portion. It has $(1-\delta)N/4$ particles in state $\ket{\psi'_2}$ that are constrained to that region of the cylinder. However, we can also find there a fraction of the particles that were successfully measured to be in state $\ket{\psi_1}$ and freely move between the leftmost portion of the gas and the central one, only constrained by the blue wall. Specifically, a fraction $(1-\delta)/(1+\delta)$ of the $(1+\delta)N/4$ particles in this situation will be found on average on the leftmost region of the cylinder. This amounts precisely to $(1-\delta)N/4$ particles in state $\ket{\psi'_1}$. Thus, the quantum state of a random particle of the gas in this region can be described by the mixed state
\begin{equation}
\hat\rho'=\frac{1}{2}\ket{\psi'_1}\!\!\bra{\psi'_1}+\frac{1}{2} \ket{\psi'_2}\!\!\bra{\psi'_2}.
\end{equation} 
The same reasoning can be carried out for the other two regions of the cylinder, with the same conclusion. 
	
Finally, before proceeding with the third step of the cycle, the demons are instructed to perform a joint resetting of the memory and the gas, with a work cost $W'_{\text{reset}}$. After this step, the memory is taken back to its default state, and the ensemble of particles in the gas can be described by
\begin{equation}
\hat\rho=\frac{1}{2}\ket{\psi_1}\!\!\bra{\psi_1} + \frac{1}{2} \ket{\psi_2}\!\!\bra{\psi_2}.
\end{equation}
As in Sec.~\ref{Section: Peres' demon}, we assume that the measurement-resetting process can be performed in a reversible way, so that
\begin{equation}\label{measurement and resetting modified}
W'_{\text{meas}} + W'_{\text{reset}}=0.
\end{equation}
	
Once the modified cycle is performed, using the work calculated in Eq.~\eqref{Eq: W'2} for the modified step 2, the condition given in Eq.~\eqref{measurement and resetting modified} for the work involved in the measurement and the resetting, and retrieving the work involved in steps 1, 3, 4, and 5 from Eqs.~\eqref{eq: W1},~\eqref{eq: W3},~\eqref{eq: W4}, and~\eqref{eq: W5}, we can calculate the net work received by the system during the cycle,
\begin{equation}
W'_t= -N k_{\textsc{b}} T \bigg[ 1 - H\bigg( \frac{1+\delta}{2} \bigg) - S(\hat\rho) \bigg] \ln 2.
\end{equation}
If we now pose the constraint that the second law is satisfied, i.e., that $W_t \geq 0$, then this translates into
\begin{equation}
1 - H\bigg( \frac{1+\delta}{2} \bigg) - S(\hat\rho) \leq 0.
\end{equation}
As $\delta$ increases from 0 to 1, $H((1+\delta)/2)$ decreases, and therefore the left-hand side of the previous inequality increases. Thus, the optimal $\delta$ that we can achieve while fulfilling the second law, which we denote $\delta_\text{th}$, is the one that saturates the inequality,
\begin{equation}\label{eq: thermodynamic bound}
H\bigg( \frac{1+\delta_\text{th}}{2} \bigg)= 1-S(\hat\rho).
\end{equation}
This is the thermodynamic bound that we were looking for. 
	
It is worth pointing out that the construction presented in this section can be extended to more general scenarios than the one considered here. In Appendix~\ref{Appendix: Generalizations of the thermodynamic bound} we outline how the cycle can be modified in order to incorporate these cases, although we will not study their associated generalized thermodynamic bounds in the present work.

	
\section{Comparison between bounds}\label{Section: Discussion}
	
\begin{figure}
\includegraphics[scale=.57]{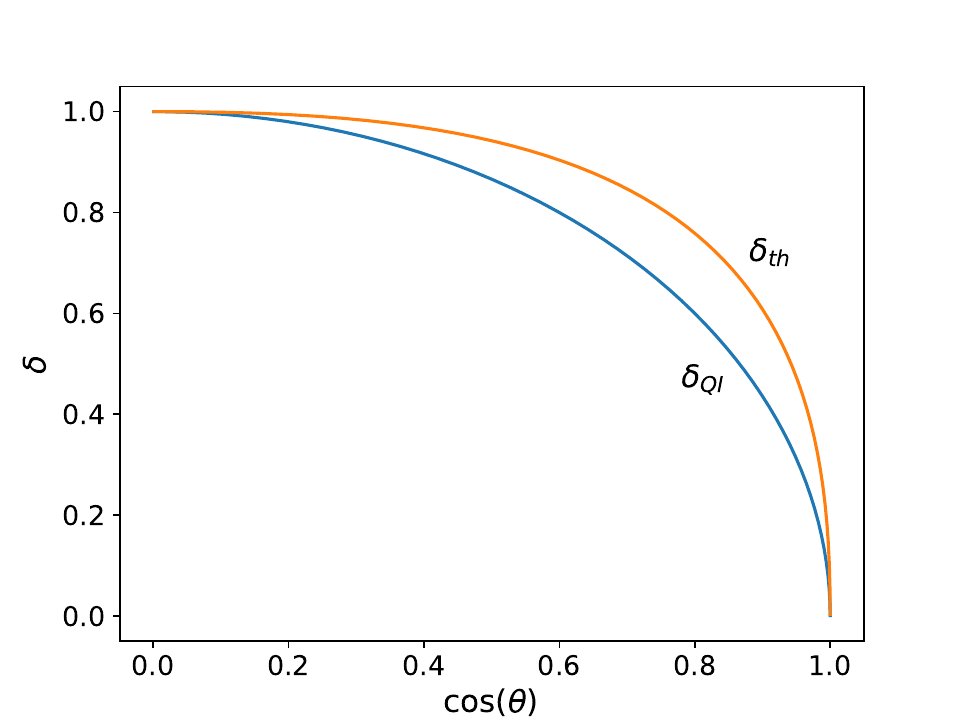}
\caption{Optimal performance for quantum state discrimination achievable within the constraints imposed by thermodynamics (orange) and quantum information theory (blue).}
\label{fig: both bounds}
\end{figure}

Now that we have obtained the thermodynamic bound, let us compare it with the bound given by the Holevo-Helstrom theorem~\cite{Holevo1974,Helstrom1976} within quantum information theory. The theorem establishes that the optimal success probability in the discrimination of two quantum states $\hat\rho_1$ and $\hat\rho_2$ is given by
\begin{equation}
p_s= \frac{1}{2} + \frac{1}{4}|\!| \hat\rho_1-\hat\rho_2 |\!|_1 \; \Rightarrow \; \delta_{\textsc{qi}}=\frac{1}{2}|\!| \hat\rho_1-\hat\rho_2 |\!|_1,
\end{equation}
where we used that $p_s=(1+\delta_\textsc{qi})/2$, with $\delta_\textsc{qi}$ representing how much better than random we can perform in the optimal case according to quantum information. For the particular case of two pure states, \mbox{$\hat\rho_1=\ket{\psi_1}\!\!\bra{\psi_1}$} and \mbox{$\hat\rho_2=\ket{\psi_2}\!\!\bra{\psi_2}$}, we can evaluate the bound in terms of the overlap between both states: let $\theta \in [0,\pi/2]$ be such that
\begin{equation}\label{Eq: def theta}
|\!\braket{\psi_1}{\psi_2}\!|=\cos \theta,
\end{equation} 
then we have (see Appendix~\ref{Appendix: results in terms of the angle} for details)
\begin{equation}
|\!| \hat\rho_1-\hat\rho_2 |\!|_1=2\sin\theta.
\end{equation}
The optimal bound that one can obtain in quantum information theory is therefore given by
\begin{equation}\label{Eq: delta_qi}
\delta_{\textsc{qi}}=\sin\theta.
\end{equation}
On the other hand, in Appendix~\ref{Appendix: results in terms of the angle} we show that
\begin{equation}
S(\hat\rho)=H\bigg( \frac{1+\cos\theta}{2} \bigg),
\end{equation}
where, as in Secs.~\ref{Section: Peres' demon} and~\ref{Section: Modified cycle and bound}, $\hat\rho$ is given by Eq.~\eqref{eq: rho}. From Eq.~\eqref{eq: thermodynamic bound},
\begin{equation}\label{eq: thermo bound explicit}
H\bigg( \frac{1+\delta_\text{th}}{2} \bigg)= 1-H\bigg( \frac{1+\cos\theta}{2} \bigg).
\end{equation}
For each \mbox{$\theta \in [0,\pi/2]$}, this equation can be solved implicitly. In Fig.~\ref{fig: both bounds}, both bounds $\delta_{\textsc{qi}}$ and $\delta_\text{th}$ are represented as functions of $\cos\theta$. 

\begin{figure}
\includegraphics[scale=.57]{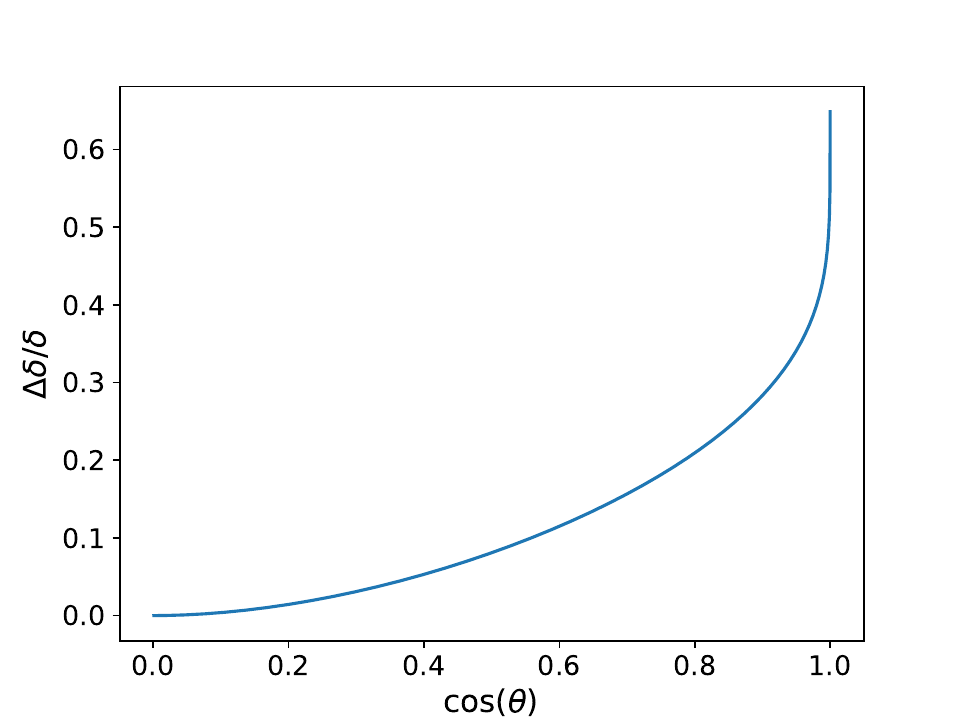}
\caption{Relative comparison between the thermodynamic and the Holevo-Helstrom bounds, \mbox{$(\delta_\text{th}-\delta_{\textsc{qi}})/\delta_\text{th}$}. It can be shown that \mbox{$(\delta_\text{th}-\delta_{\textsc{qi}})/\delta_\text{th} \to 1$} as $\cos\theta\to 1$, although this limit is not reached in the figure.}
\label{fig: relative comparison}
\end{figure} 

As expected, the thermodynamic bound is looser than the quantum information one. The relative difference $(\delta_\text{th}-\delta_{\textsc{qi}})/\delta_\text{th}$ plotted in Fig.~\ref{fig: relative comparison} reveals that when the two states considered are almost orthogonal, the thermodynamic bound is close to the (optimal) bound given by the Holevo-Helstrom theorem. As the two states get closer to each other, the thermodynamic bound stops being a good first estimation of the quantum information one, although it does show the same behaviour $\delta_\text{th} \to 0$ as $\cos\theta \to 1$. 
    
The looseness of the thermodynamic bound is revealing how far one could go if we were out of the regime of validity of quantum mechanics---since the only aspect of quantum mechanics that we used in order to derive the thermodynamic bound was its linearity. Consider, for instance, the operation performed by the demons in the measurement process. This operation is something that we assume the demons, with their supposed abilities to discriminate between $\ket{\psi_1}$ and $\ket{\psi_2}$, can perform. However, in general this operation might not be a quantum channel. As a matter of fact, whenever the assumed $\delta$ is above the quantum information limit $\delta_\textsc{qi}$ but below the thermodynamic bound $\delta_\text{th}$, this operation is indeed \textit{not a quantum channel}. As far as we know, it is an impossible operation, but it could be a possible one in some kind of beyond-quantum theory. All we are saying here is that it is forbidden by quantum mechanics, but not by thermodynamics alone.

The slackness of $\delta_\text{th}$ with respect to $\delta_\textsc{qi}$ also constitutes yet one more indicator that the von Neumann entropy is in general not appropriate for representing thermodynamic entropy~\cite{Strasberg2019,Safranek2019,Strasberg2021,Alipour2022,Adam2022}. In this regard, even if we consider different cycles and obtain different thermodynamic bounds, we would not anticipate these to be more restrictive than the quantum information bound $\delta_\textsc{qi}$. This is so because thermodynamic entropy is always coarser than the von Neumann entropy~\cite{Strasberg2021}, and therefore no process forbidden by thermodynamics should be expected to be allowed by quantum mechanics.
    
Finally, it is worth comparing the thermodynamic bound with the one we can obtain from the limit on the accessible information imposed by Holevo's bound. Specifically, suppose we only took into account the information accessed by the measurement performed by the demons, before the resetting is performed. This quantity, for a single random particle of the gas, is represented by the mutual information 
\begin{equation}\label{Eq: mutual info gas memory}
I(\text{gas}:\text{memory})=H_\text{gas} + H_\text{memory} - H_{\text{joint}},
\end{equation}
where $H_{\text{gas}}$, $H_{\text{memory}}$, and $H_{\text{joint}}$ are the Shannon entropies of the random variables described by the state of the gas particle ($\ket{\psi_1}$ or $\ket{\psi_2}$), the state of the memory (1 or 2, depending on the measurement outcome), and their joint state (formed by the state of the gas and the state measured by the demons). Since exactly half of the particles of the gas are in each quantum state, the probability distribution that describes the state of the gas particles is given by $p_{\text{gas}}(1) = p_{\text{gas}}(2) = 1/2$, and hence
\begin{equation}\label{Eq: Shannon entropy gas}
H_{\text{gas}} = H\bigg(  \frac{1}{2} \bigg) = 1.
\end{equation}
On the other hand, the probability that a memory bit registers a 1 is given by the probability that a demon measures that a random particle of the gas is in state $\ket{\psi_1}$. Since the accuracy of the demons at guessing the state of the particles is given in Eq.~\eqref{eq: prob. success} by $(1+\delta)/2$, then, by Bayes' theorem, 
\begin{equation}\label{Eq: p_memory}
p_{\text{memory}}(1) = \frac{1+\delta}{2}\, p_{\text{gas}}(1) + \frac{1-\delta}{2} \, p_{\text{gas}}(2) = \frac{1}{2},
\end{equation}
and thus $p_{\text{memory}}(2)=1/2$ as well. Therefore,
\begin{equation}
H_{\text{memory}} = H\bigg(  \frac{1}{2} \bigg) = 1.
\end{equation}
Finally, the joint distribution of the quantum state of a random gas particle and the outcome of its measurement by the demons is given by
\begin{align}
p_{\text{joint}}(j,k) & = \text{Prob}\big(\ket{\psi_j},k\big) \\
&= \frac{1+\delta}{4} \, \delta_{jk} + \frac{1-\delta}{4} \, (1 - \delta_{jk}), \nonumber
\end{align}
for $j,k \in \{1,2\}$, where $\delta_{jk}$ is the Kronecker delta. Thus,
\begin{align}
H_{\text{joint}} & = - \sum_{j,k=1}^{2} p_{\text{joint}}(j,k) \,\log_2 p_{\text{joint}}(j,k) \nonumber \\
& = - \frac{1+\delta}{2} \,\log_2 \frac{1+\delta}{4} - \frac{1-\delta}{2} \,\log_2 \frac{1-\delta}{4} \nonumber \\
& = 1 - \frac{1+\delta}{2}\,\log_2\frac{1+\delta}{2} - \frac{1-\delta}{2}\,\log_2\frac{1-\delta}{2} \nonumber \\
& = 1 + H \bigg( \frac{1+\delta}{2} \bigg).
\end{align}
From Eq.~\eqref{Eq: mutual info gas memory}, we conclude that
\begin{equation}
I(\text{gas}:\text{memory}) = 1 - H \bigg( \frac{1+\delta}{2} \bigg).
\end{equation}
Then, Holevo's theorem imposes a bound on accessible information that yields
\begin{equation}
1-H\bigg(\frac{1+\delta}{2}\bigg) \leq S(\hat\rho),
\end{equation}
which naturally leads to another quantum information bound, $\delta_{\text{Hol}}$, satisfying
\begin{equation}\label{eq: Holevo bound}
H\bigg(\frac{1+\delta_\text{Hol}}{2}\bigg) = 1 - H\bigg(\frac{1+\cos\theta}{2}\bigg).
\end{equation}
However, this implicit equation is exactly the same as Eq.~\eqref{eq: thermo bound explicit}, which defines the thermodynamic bound, and therefore \mbox{$\delta_{\text{Hol}}=\delta_\text{th}$}. This is consistent with the results of M. Plenio~\cite{Plenio1999}, who obtained Holevo's bound from Landauer's principle. It also suggests that the thermodynamic bound on accessible information obtained in~\cite{Maruyama2005} could be improved to reach Holevo's bound if one incorporates a memory resetting step in the cycle. Indeed, both the present work and~\cite{Plenio1999} suggest that Holevo's theorem can be derived from strictly thermodynamic arguments, although Holevo and Helstrom's cannot.

\section{Conclusion}\label{Section: Conclusion}
	
We have shown that thermodynamics sets a constraint on quantum state discrimination. Specifically, we established that enforcing the fulfillment of the second law of thermodynamics sets an upper limit to the accuracy with which two different pure quantum states can be distinguished. Here, we considered a specific cycle for an ideal gas whose particles have an internal quantum degree of freedom that is dynamically decoupled from their positions and momenta. We introduced two walls operated by a couple of \textit{demons} that are responsible for distinguishing the gas particles being in one state or another with some efficiency $(1+\delta)/2$. The key detail is that the work involved in some steps of the cycle depends on this efficiency. That dependency allows us to prove that whenever $\delta$ exceeds a particular \textit{thermodynamic bound}, energy can be drained from the heat bath and be completely converted into a positive amount of extracted work, in violation of the second law. This thermodynamic bound coincides with the one we can deduce from Holevo's bound on accessible information, but was found to be always looser than the one given by the Holevo-Helstrom theorem that restricts quantum state discrimination within quantum information theory.

This result has the potential to give hints on the relationship between von Neumann entropy and thermodynamic entropy~\cite{Strasberg2019,Safranek2019,Strasberg2021,Alipour2022,Adam2022}, and more specifically on how to define the latter in full generality when dealing with quantum systems. The construction used here also generalizes from pure to mixed states, and to more than two states. In particular, the thermodynamic bound resulting from the case of multiple states may give first estimations for quantum state discrimination problems that otherwise generally require solving a semidefinite program problem~\cite{Yuen1975,Bae2015,Watrous2018}. It should also be possible to modify the current setup to include entanglement, or correlations in general, between different portions of the gas. This might give some additional insight about the role that coherences play as a resource in quantum thermodynamics (see, e.g.,~\cite{Streltsov2017}).

The way the thermodynamic bound is obtained is agnostic to the specific details of the quantum formalism. The feasibility of the proposed cycle only relies on linearity and the perfect discrimination of orthogonal states, and these are therefore the only assumptions we need to make about the underlying theory. Notice also that the demons employed in the cycle are not Maxwell's demons: they have a (shared) memory, we deal with its erasure during the cycle, and they do not break any thermodynamic law. The reason why it is convenient to use this artifact is that we did not want to be specific about the physical mechanism involved in the discrimination of the states, and the so-called demons represent a suitable abstraction that allows us to focus strictly on the flow of information of the measurement-resetting steps of the cycle. The generality of the approach means that the thermodynamic bound is universal: as long as the theory governing the internal degree of freedom is linear with respect to the quantum states, it should be satisfied, even if we were considering some formalism beyond quantum mechanics that might allow for improvements on the Holevo-Helstrom bound.

 
\begin{acknowledgements}

	
The author would like to thank Luis J. Garay for their always enlightening exchanges and his meticulous criticism, Eduardo Mart\'in-Mart\'inez for his key comments on how to present the ideas of the present work, Adam Teixid\'o-Bonfill for his illuminating observations on the memory resetting process, and Shayan Majidy, T. Rick Perche, and Bruno de S. L. Torres for their thorough feedback. The author acknowledges the support of a Mike and Ophelia Lazaridis Fellowship, as well as the support of a fellowship from ``La Caixa'' Foundation (ID 100010434, code LCF/BQ/AA20/11820043). Research at Perimeter Institute is supported in part by the Government of Canada through the Department of Innovation, Science and Economic Development and by the Province of Ontario through the Ministry of Colleges and Universities.
\end{acknowledgements}
    
\onecolumngrid
\appendix

\section{Even mixture of non-orthogonal states}\label{Appendix: Even mixture of non-orthogonal states}

In this appendix, we show that as long as the states are non-orthogonal, neither of the eigenvalues of $\hat\rho$, which we denoted $c$ and $1-c$ in Eq.~\eqref{eq: rho diagonalization}, are equal to $1/2$. Equivalently, if $c$ was $1/2$, then $\ket{\psi_1}$ and $\ket{\psi_2}$ would be orthogonal. Indeed, let us assume $c=1/2$, then we would have  
\begin{align}
\bra{\psi_1} \hat\rho \ket{\psi_1}&=\frac{1}{2}\big(|\!\braket{\psi_1}{\psi_1}\!|^2+|\!\braket{\psi_1}{\psi_2}\!|^2\big)=\frac{1}{2}\big(|\!\braket{\psi_1}{\phi_1}\!|^2+|\!\braket{\psi_1}{\phi_2}\!|^2\big).
\end{align}
However, since $\ket{\phi_1}$ and $\ket{\phi_2}$ form an orthonormal basis of the subspace of the Hilbert space where $\hat\rho$ has support, and since $\ket{\psi_1}$ is a unit vector that necessarily lives in this subspace,
\begin{equation}
|\!\braket{\psi_1}{\psi_1}\!|^2=|\!\braket{\psi_1}{\phi_1}\!|^2+|\!\braket{\psi_1}{\phi_2}\!|^2=1,
\end{equation}
which implies that
\begin{equation}
\!\braket{\psi_1}{\psi_2}\!=0,
\end{equation}
as claimed.

\section{Generalizations of the thermodynamic bound}\label{Appendix: Generalizations of the thermodynamic bound}
	
In this appendix, we describe how to extend the construction leading to the thermodynamic bound presented in Sec.~\ref{Section: Modified cycle and bound} to more general scenarios.
	
\subsection{Mixed states}\label{Appendix subsection: Mixed states}
	
The first instance in which the cycle can be generalized is when the two states to be distinguished are not necessarily pure states, $\ket{\psi_1}$ and $\ket{\psi_2}$, but generic mixed states, $\rhoh_1$ and $\rhoh_2$. In this case, the cycle can be performed in the same way as before, except for two modifications. First, after performing the resetting of the memory (see Fig.~\ref{fig: modified cycle}), the state of the gas can be written as
\begin{equation}
\rhoh = \frac{1}{2}\rhoh_1 + \frac{1}{2}\rhoh_2 = \sum_{k=1}^{L} c_j \,\proj{\phi_k}{\phi_k},
\end{equation} 
for some family of orthogonal states \mbox{$\{\ket{\phi_k},\:k=1,\hdots,L\}$} that diagonalize $\rhoh$. Notice that unlike in Eq.~\eqref{eq: rho diagonalization}, here the diagonal form of $\rhoh$ will in general require more than two terms (i.e., $L>2$). Thus, in this case the third step would have to be modified to include a sequence of separations, instead of a single one. Specifically, we can first separate $\ket{\phi_1}$ from the rest, so that in a fraction $1/L$ of the volume of the cylinder the state of the gas is $\proj{\phi_1}{\phi_1}$, while in the remaining portion the state can be written as
\begin{equation}
\rhoh' = \frac{1}{1-c_1}\sum_{j=2}^L c_j \proj{\phi_j}{\phi_j}. 
\end{equation}
The operation can be repeated in this remaining fraction of the cylinder, separating $\ket{\phi_2}$ from the rest. After $L$ separations, the cylinder is divided into $L$ equal portions of gas, each of them in one of the states $\ket{\phi_k}$, $k=1,\hdots,L$. 

The second modification concerns the last step of the cycle, that takes the system back to the initial state where $\rhoh_1$ and $\rhoh_2$ are each occupying a quarter of the volume of the cylinder. Since $\rhoh_1$ and $\rhoh_2$ are now generic mixed states, and not pure states, the unitary operations applied originally in step 5 (see Fig.~\ref{fig: modified cycle}) are not enough to recover the initial state in this case. Instead, taking into account that $\rhoh_1$ and $\rhoh_2$ can be diagonalized as 
\begin{equation}
\rhoh_1 = \sum_{i=1}^{N} a_i \proj{\psi_{1i}}{\psi_{1i}}, \quad \text{and} \quad \rhoh_2 = \sum_{j=1}^{M} b_{j} \proj{\psi_{2j}}{\psi_{2j}},
\end{equation}
the original unitaries of step 5 can be modified so that they take the appropriate portions of the gas to each one of the states $\ket{\psi_{1i}}$ and $\ket{\psi_{2j}}$. An additional final step is necessary to complete the cycle: the different portions of the gas that correspond to each one of the states $\rhoh_1$ and $\rhoh_2$ need to be mixed. Notice that this mixing has an associated increase of entropy that has to be taken into account to obtain the thermodynamic bound for this case.

		
\subsection{Unequally likely states}\label{Appendix subsection: Unequally likely states}
	
The second generalization of the cycle comes from observing that, in a generic quantum state discrimination problem, the two states to be identified do not necessarily have equal probabilities of appearance. Specifically, let us denote with $\gamma$ the probability with which the system given to the ``guesser'' is in state $\rhoh_1$, so that $1-\gamma$ is the probability that the system is in state $\rhoh_2$ instead. For our setup, this translates into
\begin{equation}
p_{\text{gas}}(1) = \gamma, \quad \text{and} \quad p_{\text{gas}}(2)=1-\gamma.
\end{equation}
The cycle analyzed in Sec.~\ref{Section: Modified cycle and bound} corresponds to the special case $\gamma=1/2$ (cf. Eqs.~\eqref{Eq: Shannon entropy gas} and~\eqref{Eq: p_memory}), i.e., when both states are equally likely, but it can be easily modified to include the more general case in which $\gamma$ is any number in $(0,1)$. Namely, it suffices to start the cycle with the ideal gas divided in two portions of $\gamma N$ and $(1-\gamma)N$ particles whose internal degrees of freedom are in states $\rhoh_1$ and $\rhoh_2$, each of them initially occupying fractions $\gamma/2$ and $(1-\gamma)/2$ of the total volume of the cylinder, respectively. The rest of the cycle follows in a completely analogous way.

\subsection{More than two states}\label{Appendix subsection: More than two states}

The last level of generalization consists of extending the analysis to the case where the collection of states $\{\hat\sigma_1,\hdots,\hat\sigma_n\}$ that we aim to discriminate has more than two elements, i.e., when $n\geq 3$. Specifically, the goal is to maximize the probability of success
\begin{equation}\label{Eq: delta_n}
p_{s} = \sum_{j=1}^n \operatorname{Prob}(\text{``guessing correctly }j \text{''})\, p(j)=\frac{1+\delta_n}{n},
\end{equation}
where, as in Appendix~\ref{Appendix subsection: Unequally likely states}, $p(j)$ is the probability with which the ``guesser'' is given the system in state $\rhoh_j$. Eq.~\eqref{Eq: delta_n} is the $n$-states analog of Eq.~\eqref{eq: prob. success}. Bounding $p_s$, as for the $n=2$ case, amounts to finding an upper bound for $\delta_n$, which is a measure of how much better than random guessing we can perform. It turns out that a construction to derive a thermodynamic bound for $\delta_n$ can be obtained using a combination of the two previous generalizations. Specifically, for each $k\in\{1,\hdots,n\}$, we consider the performance of a cycle where
\begin{equation}
\rhoh_1 = \hat\sigma_k, \quad \text{and} \quad \rhoh_2 = \frac{1}{1-p(k)} \sum_{j \neq k} p(j) \,\hat\sigma_k,
\end{equation}
and
\begin{equation}
p_{\text{gas}}(1)=\gamma=p(k).
\end{equation}
Imposing the fulfillment of the second law in \textit{every one} of these $n$ cycles---assuming the same demons are used in all of them---gives a (thermodynamic) bound for $\delta_n$, as desired.

It is worth noticing that there is not necessarily a unique way to generalize the construction given in Sec.~\ref{Section: Modified cycle and bound} to more than two states (e.g., one could come up with new geometries to implement an interface between the $n$ portions of gas). However, the one given has the advantage of its simplicity and the fact that it does not require a significant alteration of the original setup.

\section{Results in terms of the angle $\theta$}\label{Appendix: results in terms of the angle}

In this appendix, for the sake of completeness, we show the derivation of two simple results used in Sec.~\ref{Section: Discussion} to compare the thermodynamic bound with the quantum information ones. 

Let $\hat\rho_1=\ket{\psi_1}\!\!\bra{\psi_1}$ and $\hat\rho_{2}=\ket{\psi_2}\!\!\bra{\psi_2}$ be two pure states. We can always write
\begin{equation}
\braket{\psi_1}{\psi_2}=e^{\ii\phi}\cos{\theta},
\end{equation}
for some $\phi\in[0,2\pi)$ and $\theta \in [0,\pi/2]$. Consider first the case in which $\theta \neq 0$. Then, 
\begin{equation}
\ket{\psi_2}=e^{\ii\phi} \cos\theta \ket{\psi_1} + \sin\theta \ket{\varphi},
\end{equation}
where
\begin{equation}
\ket{\varphi}=\frac{1}{\sin\theta} (\ket{\psi_2}-\braket{\psi_1}{\psi_2}\ket{\psi_1})
\end{equation}
has unit norm and is orthogonal to $\ket{\psi_1}$. Therefore, we can write
\vspace{2mm}
\begin{align}
\hat\rho_2&=\cos^2\theta \ket{\psi_1}\!\!\bra{\psi_1} + e^{\ii\phi} \cos\theta \sin\theta \ket{\psi_1}\!\!\bra{\varphi} + e^{-\ii\phi} \cos\theta \sin\theta \ket{\varphi}\!\!\bra{\psi_1} + \sin^2\theta \ket{\varphi}\!\!\bra{\varphi}.
\end{align}
Thus,
\begin{align}
\hat\rho_1-\hat\rho_2&=(1-\cos^2\theta) \ket{\psi_1}\!\!\bra{\psi_1} - e^{\ii\phi} \cos\theta \sin\theta \ket{\psi_1}\!\!\bra{\varphi} - e^{-\ii\phi} \cos\theta \sin\theta \ket{\varphi}\!\!\bra{\psi_1} - \sin^2\theta \ket{\varphi}\!\!\bra{\varphi},
\end{align}
and since $\ket{\psi_1}$ and $\ket{\varphi}$ are orthogonal, the eigenvalues of this operator are just the roots of the characteristic polynomial
\begin{align}
p_{-}(\lambda)&=\left| \begin{array}{cc}
1-\cos^2\theta - \lambda & -e^{\ii\phi}\cos\theta\sin\theta \\
-e^{-\ii\phi}\cos\theta\sin\theta & -\sin^2\theta - \lambda \\
\end{array} \right| =\lambda^2-\sin^2\theta,
\end{align}
which are
\begin{equation}
\lambda_{-}^{(\pm)}=\pm\sin\theta.
\end{equation}
Thus,
\begin{equation}\label{Eq: trace norm difference}
|\!|\hat\rho_1-\hat\rho_2|\!|_1=|\lambda_{-}^{(+)}|+|\lambda_{-}^{(-)}|=2\sin\theta.
\end{equation}
Similarly,
\begin{align}
\hat\rho&=\frac{1}{2}(\hat\rho_1+\hat\rho_2) =\frac{1+\cos^2\theta}{2}\ket{\psi_1}\!\!\bra{\psi_1} + \frac{e^{\ii\phi}}{2}\cos\theta\sin\theta\ket{\psi_1}\!\!\bra{\varphi} + \frac{{e}^{-\ii\phi}}{2}\cos\theta\sin\theta\ket{\varphi}\!\!\bra{\psi_1}+\frac{\sin^2\theta}{2}\ket{\varphi}\!\!\bra{\varphi}.
\end{align}
The eigenvalues of $\hat\rho$ are then the roots of the characteristic polynomial
\begin{align}
p_{+}(\lambda)&=\left| \begin{array}{cc}
\frac{1+\cos^2\theta}{2} - \lambda & \frac{e^{\ii\phi}}{2}\cos\theta\sin\theta \\
\frac{e^{-\ii\phi}}{2}\cos\theta\sin\theta & \frac{\sin^2\theta}{2} -\lambda \\
\end{array} \right| =\lambda^2-\lambda+\frac{\sin^2\theta}{4},
\end{align}
namely,
\begin{equation}
\lambda_{+}^{(\pm)}=\frac{1\pm\cos\theta}{2}.
\end{equation}
We conclude that
\begin{align}\label{Eq: entropy rho}
S(\hat\rho)&=-\lambda_{+}^{(+)}\log_2\lambda_{+}^{(+)}-\lambda_{+}^{(-)}\log_2\lambda_{+}^{(-)}=H\bigg( \frac{1+\cos\theta}{2}\bigg).
\end{align}
The case $\theta=0$ is dealt with easily, since in that case $\rhoh_1-\rhoh_2=0$ and thus
\begin{equation}
|\!|\hat\rho_1-\hat\rho_2|\!|_1 = 0,
\end{equation}
which coincides with the application of Eq.~\eqref{Eq: trace norm difference} when $\theta=0$. Similarly, in this case $\rhoh=\rhoh_1=\rhoh_2$ is a pure state, which implies that
\begin{equation}
S(\rhoh)=0,
\end{equation}
in agreement with Eq.~\eqref{Eq: entropy rho}, since $H(1)=0$.

\twocolumngrid	
\bibliography{references}
 
\end{document}